\documentclass{article}
\usepackage{ijcai18}

\usepackage{times}
\usepackage{xcolor}
\usepackage{soul}
\usepackage[utf8]{inputenc}
\usepackage[small]{caption}

\usepackage{amsmath}
\usepackage{amsfonts}
\usepackage{amsthm}
\usepackage{mathrsfs}
\usepackage{graphicx}
\usepackage{booktabs}
\usepackage{framed}

 \usepackage{colortbl}
 \usepackage{booktabs}

\title{Payoff Control in the Iterated Prisoner's Dilemma}

\author{
Dong Hao$^1$,
Kai Li$^2$ \and Tao Zhou$^1$
\\
$^1$ University of Electronic Science and Technology of China, Chengdu, China\\
$^2$ Shanghai Jiao Tong University, Shanghai, China\\
haodong@uestc.edu.cn,
kai.li@sjtu.edu.cn,
zhutou@ustc.edu
}
\begin{document}

\maketitle
\begin{abstract}
%Understanding what is the best strategy for intelligent agents in a long-run relationship is a fundamental challenge in many disciplines. Repeated game has long been the touchstone model for long-run relationships. Previous results suggest it is of particular difficulty for a repeated game player to exert an autocratic control on both players' payoffs. In this work we discover that, the scale of a single player's capability to unilaterally influence the payoffs may have been much underestimated. Under the conventional iterated prisoner's dilemma, we develop a general framework for controlling the feasible region in which the payoff pairs lie. With this framework, many well-known existing strategies can be categorized and various new payoff control strategies can be further identified. Specifically, we show that a single player can unilaterally restrict two players' payoff pairs to her objective region, thereby freely set the boundaries of the opponent's payoff, or ensure herself a wining outcome, or enforce the game to evolve to a mutual cooperation equilibrium.
Repeated game has long been the touchstone model for agents' long-run relationships. Previous results suggest that it is particularly difficult for a repeated game player to exert an autocratic control on the payoffs since they are jointly determined by all participants. This work discovers that the scale of a player's capability to unilaterally influence the payoffs may have been much underestimated. Under the conventional iterated prisoner's dilemma, we develop a general framework for controlling the feasible region where the players' payoff pairs lie. A control strategy player is able to confine the payoff pairs in her objective region, as long as this region has feasible linear boundaries. With this framework, many well-known existing strategies can be categorized and various new strategies with nice properties can be further identified. We show that the control strategies perform well either in a tournament or against a human-like opponent.

%Specifically, we show that a controller can freely determine the maximum and minimum values of the opponent's payoff function, or ensure herself a winning outcome no matter what the opponent's strategy is, or enforce the game to evolve to a mutual cooperation equilibrium.

%We propose a general framework for payoff control strategies (PCS) in repeated games, by following the classic iterated prisoner's dilemma model. A PCS player to control the payoff of his opponent. Facing a rational opponent, a PCS player can realize (i) payoff-control: set the upper bound, lower bound and even the whole region of the possible payoffs, or (ii) cooperation-enforcing: enforce the game to converge to a mutual-cooperation situation, or (iii) winning: alternatively

\end{abstract}
\section{Introduction}
Understanding what are the best strategies for intelligent agents in a long-run relationship is a fundamental challenge in many disciplines. Repeated games are prevailing tools for modeling and analyzing intelligent agents' long-run relationships \cite{mailath2006repeated}, which have been richly studied in economics, artificial intelligence and biology \cite{kandori2002introduction,claus1998dynamics,nowak1995automata}. For multi-agent systems, repeated games are widely utilized for understanding how cooperation or competition emerges among agents and how cooperative or winning strategies can be identified. It has been commonly accepted that in such games, it is impossible for a unilateral player to freely control the payoffs and determine the evolutionary route of the game, since the outcomes are jointly determined by all participants.

In this paper, we propose a general framework for payoff control in iterated prisoner's dilemma, which is a conventional model for repeated games. First of all, based on the game's Markov decision process (MDP), the correlation between a single player's strategy and the MDP's joint stationary distribution is derived. Then according to this correlation, we establish a general payoff control framework, under which a control strategy can be easily obtained by solving a system of linear inequalities. Using the payoff control framework, as long as the control objective is feasible, a controller can restrict the relation between her and the opponent's payoffs (represented by a two-tuple) to an arbitrary region with linear boundaries. More specifically, she can (i) unilaterally determine the maximum and minimum values of the opponent's possible payoffs; or (ii) always win the game no matter what the opponent's strategy is, and she can even control her winning probability; or (iii) control the evolutionary route of the game, as long as the opponent is rational and self-optimizing, the controller can enforce the game to finally converge either to a mutual-cooperation equilibrium or to any feasible equilibrium that she wishes. We simulate serval specific strategies generated under the payoff control framework in a tournament similar to that of Axelrod \cite{axelrod1981evolution}, it is found that the new payoff control strategies have remarkably good performances.

%To the best of our knowledge, this is the first general framework for a free payoff control in repeated games.

%, while the classic memory-one strategies and the existing payoff control methods can be covered as special cases under this framework.

%Moreover, this framework is simple to use, since generating a control strategy with specific objective can be done via solving a basic linear programming problem.

The discussion of payoff control in games can be traced back to \cite{boerlijst1997equal}, in which the authors discovered that, in iterated prisoner's dilemma, one player can set the opponent's payoff to a certain value. However, what is the underlying mechanism for such strategies to exist and  how to formally derive them are not thoroughly investigated. In recent years, Press and Dyson's discovery of ``zero-determinant (ZD)'' strategies illuminates a new starting point for the control \cite{dong2014zero}. They show that in repeated games, it is possible for a player to unilaterally enforce a \emph{linear} relation between her and the opponent's payoff \cite{press2012iterated}. This is the first time that the linear payoff relation control is formally investigated, which receives a lot of attention. Thereafter, the linear control on players' payoff relations is discovered in multiplayer games \cite{pan2015zero,hilbe2014cooperation}, games with imperfect information \cite{chen2014robustness,hao2015extortion} and evolutionary games \cite{adami2013evolutionary,hilbe2013evolution}. Furthermore, from a mathematical point of view, Akin formally investigated why such linear control exists in games which can be represented by an MDP and proposed a new payoff control scheme whereby one player can fix the upper bound of the opponent's payoff to the mutual cooperation reward $R$, and such strategies can enforce the game to converge to a mutual cooperation situation \cite{akin2012stable}. Extended cases of Akin's cooperation-enforcing control are then studied and special cases of the nonlinear payoff relation control are identified \cite{hilbe2015partners}.

These existing payoff control strategies confront two major problems. The first one is that they only realize special cases of payoff control such as linear control or cooperative control. Zero-determinant based strategies can only realize linear payoff relations, which are very strong and sometimes not easy to use; Akin's method based strategies can only control the upper bound of the opponent's payoff to a mutual-cooperation reward $R$. However, how to establish a general control framework with multiple and free control objectives is still challenging. The second problem is that these strategies are mostly difficult to obtain. For the zero-determinant based strategies, calculating the determinant of a matrix already has high computational complexity; for strategies based on Akin's method, when one tries to add more objectives other than cooperation-enforcement, the computational complexity increases exponentially and deriving the strategy becomes intractable. In this paper, we propose a general payoff control framework which conquers both of these two problems. In section 2, the repeated game is modeled as an MDP and the relationship between a single player's strategy and the stationary distribution of the MDP is derived. In section 3, we realize a control on the opponent's maximum and minimum payoffs. In section 4, this is extended to a free regional payoff control with multiple linear boundaries, and various types of regional control strategies, especially the cooperation-enforcing control strategies, are identified. To analyze the performances of the payoff control strategies when confronting various famous strategies, in section 5, we simulate control strategies in the Axelrod’s tournament. In the last section, to evaluate how payoff control strategies perform in the real world, we simulate them against a reinforcement learning player \cite{sutton1998reinforcement}.

\section{Strategy and Game Convergence}
The iterated prisoner's dilemma (IPD) is the canonical example for analyzing the cooperation and competition in agents' long-run relationships. The IPD consists of multiple rounds of stage games. In each stage, player $i \in \{X,Y\}$ adopts an action $a_i \in  \{C,D\}$ with a certain probability, where $C$ denotes cooperation and $D$ denotes defection. The space of the outcomes in each stage game is $\Omega=\{CC,CD,DC,DD\}$. If both players cooperate ($CC$), then each earns a reward $R$; if one cooperates but the other defects ($CD$ or $DC$), then the cooperator earns $S$ and the defector earns $T$; if they both defect ($DD$), then both get $P$. The payoff vector of player X over $\Omega$ is thus defined as ${\bf S}_X=(R,S,T,P)$ and for player Y it is ${\bf S}_Y=(R,T,S,P)$. In this paper we consider the case that player X chooses her action conditioning only on the outcome of the previous stage. It is worth noting that, in infinitely repeated games it has been proved that such one-stage memory strategies have no disadvantages as if the opponent has a longer memory \cite{press2012iterated}. The strategy of player X is defined as a vector of probabilities ${\bf p}=(p_1, p_2, p_3, p_4)$, where each component is a probability that she cooperates with player Y conditioning on the last stage outcomes $CC$, $CD$, $DC$ or $DD$, respectively. Analogously, the strategy of player Y is a vector of probabilities for cooperation ${\bf q}=(q_1,q_2, q_3,q_4)$ conditioning on the previous outcomes $CC$, $DC$, $CD$ or $DD$, respectively.

Then the transition matrix over the state space between adjacent stage games is derived as $\bf M$:
{\small
\begin{eqnarray}\label{transition_matrix}
\left[ {\begin{array}{*{5}c}
   {p_1 q_1 }  \\
   {p_2 q_3 }  \\
   {p_3 q_2 }  \\
   {p_4 q_4 }  \\
\end{array}\begin{array}{*{5}c}

\end{array}\begin{array}{*{5}c}
   {p_1 \left( {1 - q_1 } \right)}  \\
   {p_2 \left( {1 - q_3 } \right)}  \\
   {p_3 \left( {1 - q_2 } \right)}  \\
   {p_4 \left( {1 - q_4 } \right)}  \\
\end{array}\begin{array}{*{5}c}

\end{array}\begin{array}{*{5}c}
   {\left( {1 - p_1 } \right)q_1 }   & {\left( {1 - p_1 } \right)\left( {1 - q_1 } \right)} \\
   {\left( {1 - p_2 } \right)q_3 }   & {\left( {1 - p_2 } \right)\left( {1 - q_3 } \right)} \\
   {\left( {1 - p_3 } \right)q_2 }   & {\left( {1 - p_3 } \right)\left( {1 - q_2 } \right)} \\
   {\left( {1 - p_4 } \right)q_4 }   & {\left( {1 - p_4 } \right)\left( {1 - q_4 } \right)}\\
\end{array}} \right]
\end{eqnarray}
} If this Markov matrix is regular, it has the unique stationary distribution ${\bf v}=(v_1,v_2,v_3,v_4)$, which is a probability distribution over the state space $\Omega$ and can be calculated as
\begin{eqnarray}
\mathbf{v}=\lim_{n\rightarrow \infty } \frac{1}{n}\sum_{t=1}^{n}\mathbf{v}^t,
\end{eqnarray}
where ${\mathbf v}^t$ is the distribution over $\Omega$ at the $t$-th stage. Then the average expected payoffs for players X and Y are derived as $
s_X={\bf v} \cdot {\bf S}_X=(v_1,v_2,v_3,v_4) \cdot (R,S,T,P)
$ and $s_Y={\bf v} \cdot {\bf S}_Y=(v_1,v_2,v_3,v_4) \cdot (R,T,S,P)$, respectively. In the $t$-th stage, the total probability that player X cooperates is $p_c^{t}=\left (1,1,0,0 \right )\cdot {\bf v}^{t}$. And the probability she will cooperate in the next stage game is calculated as $p_c^{t+1}={\bf p}\cdot{\bf v}^{t}$. Deriving the difference between these two probabilities, we have:
\begin{eqnarray}\label{sump}
p_c^{t+1}-p_c^{t}=\left ( {\bf p}-\left ( 1,1,0,0 \right ) \right )\cdot {\bf v}^{t}.
\end{eqnarray}
Denote $\widetilde{\bf p}={\bf p}-\left ( 1,1,0,0 \right )$. Essentially, this vector depicts to what extend of speed player X changes its probability for cooperation. Sum eq. (\ref{sump}) from $t=1$ to $t=n$, then we have
\begin{eqnarray}
\sum_{t=1}^{n}{\widetilde{\bf p}} \cdot {\bf v}^{t}=\sum_{t=1}^{n}p_c^{t+1}-p_c^{t} = p_c^{n+1} - p_c^{1}.
\end{eqnarray}
If the game is infinitely repeated, averaging the above equation when $n \to \infty$ ensures that
\begin{eqnarray}
\widetilde{\bf{p}} \cdot {\bf{v}} = \mathop {\lim }\limits_{n \to \infty } \frac{1}{n} \sum_{t=1}^{n} {\widetilde{\bf{p}} \cdot {{\bf{v}}^t} }= \mathop {\lim }\limits_{n \to \infty } \frac{1}{n}\left( {p_c^{n+1} - p_c^1} \right) = 0,
\end{eqnarray}
where $\bf v$ is just the stationary distribution of the game's state transition matrix. Expanding the vector equation $\widetilde{\bf{p}} \cdot {\bf{v}}=0$ leads to:
\begin{eqnarray}\label{expanded_vector_equation2}
\left( { - 1 + {p_1}} \right){v_1} + \left( { - 1 + {p_2}} \right){v_2} + {p_3}{v_3} + {p_4}{v_4} = 0.
\end{eqnarray}
This relation is firstly discovered in \cite{press2012iterated} and then investigated in \cite{akin2012stable}. It significantly reveals the underlying relationship between the game's transition matrix and the unilateral strategy of a single player.

\section{Control the Maximum and Minimum Values of Opponent's Payoff}
%\subsubsection{Upper bound of player Y's payoff}
If player X's objective is to ensure that Y's expected payoff
\begin{eqnarray}\label{upper_bound_inequation1}
{s_Y} \le W,
\end{eqnarray}
where $
s_Y={\bf v} \cdot {\bf S}_Y=(v_1,v_2,v_3,v_4) \cdot (R,T,S,P)
$ and $W$ is a constant,
then the objective function eq. ($\ref{upper_bound_inequation1}$) becomes
${\left( {{v_1},{v_2},{v_3},{v_4}} \right) \cdot ( \left( {R,T,S,P} \right) - (W,W,W,W)} ) \le 0$. Multiplying both side with a positive factor $1-p_2$ does not change the inequality, thus eq. ($\ref{upper_bound_inequation1}$) is equivalent to $
\left( {{s_Y} - W} \right)\left( {1 - {p_2}} \right) \le 0 $.
Substituting eq. (\ref{expanded_vector_equation2}) into it and combining the coefficients of $v_i$, the objective of player X is finally reduced to an inequality as follows.
\begin{eqnarray}\label{constrain_for_upper_bound}
{{\alpha _1}{v_1} + {\alpha _3}{v_3} + {\alpha _4}{v_4} \le 0},
\end{eqnarray}
%
%satisfying the constrains for controlling the upper bound of the opponent's payoff is equivalent to finding the solution of the set bellow:
%\begin{eqnarray}\label{constrain_for_upper_bound}
%\left\{ {{\alpha _1},{\alpha _3},{\alpha _4}|\forall {v_1},{v_3},{v_4}\left( {{\alpha _1}{v_1} + {\alpha _3}{v_3} + {\alpha _4}{v_4} \le 0} \right)} \right\},
%\end{eqnarray}
where $\alpha _1$, $\alpha _3$ and $\alpha _4$ are the coefficients and
\begin{eqnarray}\label{constrain_for_upper_bound_alphas}
\left\{ \begin{array}{l}
 {\alpha _1} = \left( {R - W} \right)\left( {1 - {p_2}} \right) + (W - T)(1 - {p_1}), \\
 {\alpha _3} = (T - W){p_3} + (S - W)(1 - {p_2}), \\
 {\alpha _4} = (T - W){p_4} + (P - W)(1 - {p_2}). \\
 \end{array} \right.
 \end{eqnarray}
One sufficient condition for eq. (\ref{constrain_for_upper_bound}) to hold is that all $\alpha_i$ are non-positive. This further requires that the strategy of player X should fall into the following region:
\begin{eqnarray}\label{constrain_for_upper_bound_final}
\left\{ \begin{array}{l}
 0 \le {p_2} < 1, \\
 0 \le {p_1} \le \min \left( {1 - \frac{{R - W}}{{T - W}}(1 - {p_2}),1} \right), \\
 0 \le {p_3} \le \min \left( {\frac{{W - S}}{{T - W}}(1 - {p_2}),1} \right), \\
 0 \le {p_4} \le \min \left( {\frac{{W - P}}{{T - W}}(1 - {p_2}),1} \right). \\
 \end{array} \right.
\end{eqnarray}

It is shown that there are multiple candidate strategies for player X to control the maximum value of Y's possible payoff. Moreover, how to choose $p_1, p_3$ and $p_4$ depends on the value of $p_2$, which is the probability that X will cooperate after she cooperated but was defected by the opponent. The value of $p_2$ partially reflects X's bottom line for cooperation and all the other components in $\bf p$ should be adjusted accordingly. Nevertheless, as long as eqs. (\ref{constrain_for_upper_bound_final}) has solutions, no matter which level of such a bottom line player X has, it is always possible for her to control the maximum value of player Y's payoff to her objective $W$.

%We illustrate the effect of some Nash strategies of player X in the following fig. (1). It is shown that by using this strategy $\bf p$, the controller X set the upper bound of Y's payoff as $W=1.5$ The black points show the payoff pairs while the red line shows the upper bound. One can see the payoff pairs never exceed the redline and Y's payoff is successfully suppressed under $1.5$. It's worth noting that, such a payoff controlling strategy $\bf p$ is only one solution of Eqs. \ref{constrain_for_upper_bound_final}. As long as Eqs. \ref{constrain_for_upper_bound_final} can be satisfied with $W=1.5$, controller X can always control the upper bound of Y's payoff as $1.5$, while the shape of the region of the payoff pairs varies.

%\begin{figure}[h]\label{fig_only_upper_bound}
%    \centering
%	\includegraphics[width=0.23\textwidth]{figure/upperbound_W1.5_P08-04-10-00.eps}
%    \includegraphics[width=0.23\textwidth]{figure/lowerbound_W0.5_P10-00-06-02.eps}
%    \caption{(A) X's strategy ${{\bf{p}}} = \left( {0.8,0.4,1.0,0} \right)$ is derived according to eqs. (\ref{constrain_for_upper_bound_final}) with the objective $s_Y \le W$ and $W=1.5$ and Y's strategy $\bf q$ is randomly sampled 10000 times. (B) X's strategy ${{\bf{p}}} = \left( {1.0,0,0.6,0.2} \right)$ is derived according to eqs. (\ref{constrain_for_lower_bound_final}) with the objective $s_Y\ge U$ and $U=0.5$ and Y's strategy $\bf q$ is randomly sampled 10000 times.}
%\end{figure}

Based on the above model, besides controlling the maximum value of Y's payoff, player X can also control the the minimum payoff Y can achieve. If X's objective is:
\begin{eqnarray}\label{constrain_for_lower_bound_imply}
{s_Y} \ge U,
\end{eqnarray}
then X actually secures a bottom line for Y's payoff. Applying the similar trick as for solving eq. (\ref{upper_bound_inequation1}), the constraints for eq. ($\ref{constrain_for_lower_bound_imply}$) becomes:
%\begin{eqnarray}\label{constrain_for_lower_bound_imply_to_inequation}
%\begin{array}{l}
%\left[ {\left( {{v_1},{v_2},{v_3},{v_4}} \right) \cdot \left( {R,S,T,P} \right) - W} \right]\left( {1 - {p_1}} \right) \ge 0 \\
% \end{array}
%\end{eqnarray}
%Still, substituting Eq. (\ref{expanded_vector_equation2}) into Eqs. (\ref{constrain_for_lower_bound_imply_to_inequation}) leads to a problem of finding the solution of the following set problem:
\begin{eqnarray}\label{constrain_for_lower_bound_set}
{{\beta _1}{v_1} + {\beta _3}{v_3} + {\beta _4}{v_4} \ge 0} ,
\end{eqnarray}
where
\begin{eqnarray}\label{constrain_for_lower_bound_inequation_set}
\left\{ \begin{array}{l}
 {\beta _1} =  \left( {R - U} \right)\left( {1 - {p_2}} \right) + \left( {U - T} \right)\left( {1 - {p_1}} \right), \\
 {\beta _3} = \left( {T - U} \right){p_3} + \left( {S - U} \right) \left( {1 - {p_2}} \right), \\
 {\beta _4} = \left( {T - U} \right){p_4} + \left( {P - U} \right) \left( {1 - {p_2}} \right). \\
 \end{array} \right.
\end{eqnarray}
Thus one sufficient condition for ${s_Y} \ge U$ is that all $\beta_i$ are non-negative, then the solution is:
\begin{eqnarray}\label{constrain_for_lower_bound_final}
\left\{ \begin{array}{l}
 0 \le {p_2} < 1, \\
 \max \left( {0,{\kern 1pt} {\kern 1pt} {\kern 1pt} 1 - \frac{{R - U}}{{T - U}}(1 - {p_2})} \right) \le {p_1} \le 1, \\
 \max \left( {0,{\kern 1pt} {\kern 1pt} {\kern 1pt} \frac{{U - S}}{{T - U}}(1 - {p_2})} \right) \le {p_3} \le 1, \\
 \max \left( {0,{\kern 1pt} {\kern 1pt} {\kern 1pt} \frac{{U - P}}{{T - U}}(1 - {p_2})} \right) \le {p_4} \le 1. \\
 \end{array} \right.
 \end{eqnarray}

%We show an example for controlling player Y's lower bound in Figure (2). By using this strategy $\bf p$, the controller X set the lower bound of Y's payoff as $U=0.5$ The black points show the payoff pairs while the red line shows the upper bound. One can see the payoff pairs never exceed the redline and Y's payoff is successfully suppressed under $0.5$. It's worth noting that, such a payoff controlling strategy $\bf p$ is only one solution of Eqs. \ref{constrain_for_lower_bound_final}. As long as Eqs. \ref{constrain_for_lower_bound_final} can be satisfied with $U=0.5$, controller X can always control the upper bound of Y's payoff as $0.5$, while the shape of the region of the payoff pairs varies. It is worth noting that, the setting of this controlling strategy grantees that mutual cooperation is the optimal strategy for both players, since the region of the payoff pairs illustrates that the maximum payoff for both players are $R$, which is generated by the both players cooperating with each other. A controlling strategy leading to a payoff pair region with such a shape, i.e., with the point $\left(R,R\right)$ being an extreme point in the convex hull.
%\begin{figure}[h]\label{fig_only_lower_bound}
%    \centering
%	\includegraphics[width=0.3\textwidth]{figure/lowerbound_W0.5_P10-00-06-02.eps}
%    \caption{X's strategy ${{\bf{p}}} = \left( {1.0,0,0.6,0.2} \right)$ is derived according to eqs. (\ref{constrain_for_upper_bound_final}) with the objective $s_Y\ge U$ and $U=0.5$ and Y's strategy $\bf q$ is randomly sampled 10000 times.}
%\end{figure}
\begin{figure}\label{fig_same_upper_bound_different_strategies}
    \centering
	\includegraphics[width=0.15\textwidth]{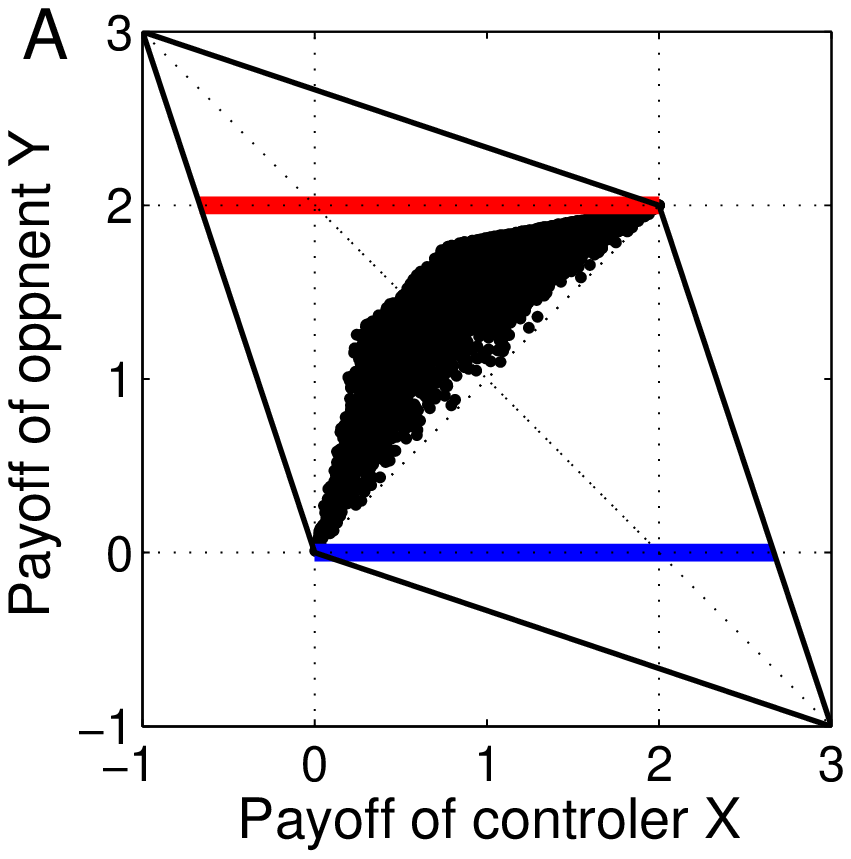}
    \includegraphics[width=0.15\textwidth]{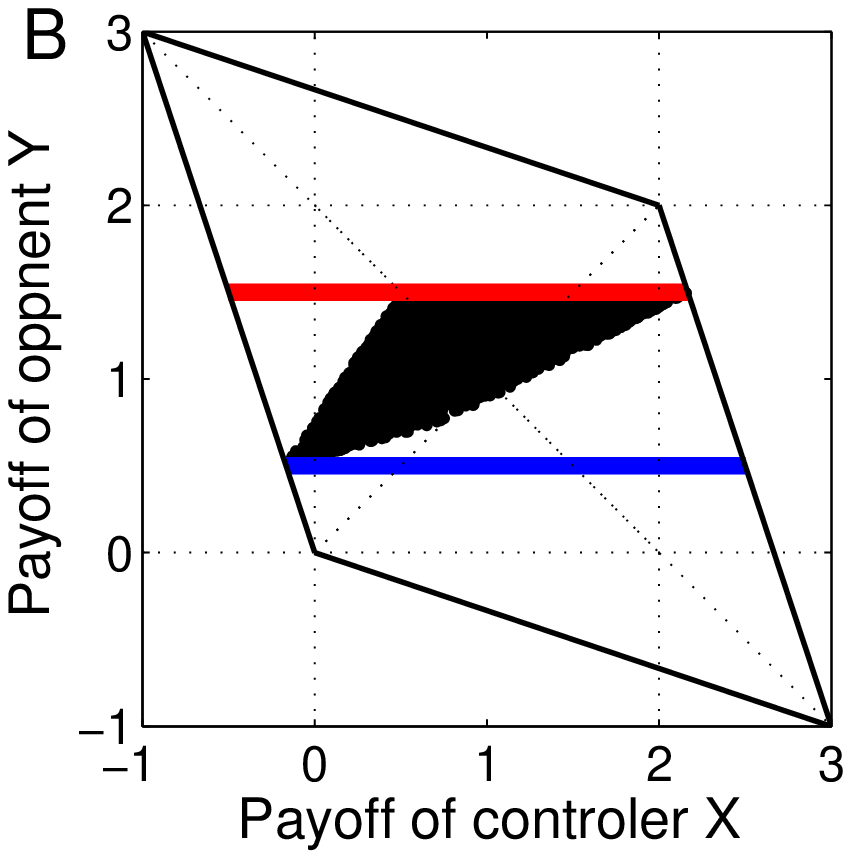}
    \includegraphics[width=0.15\textwidth]{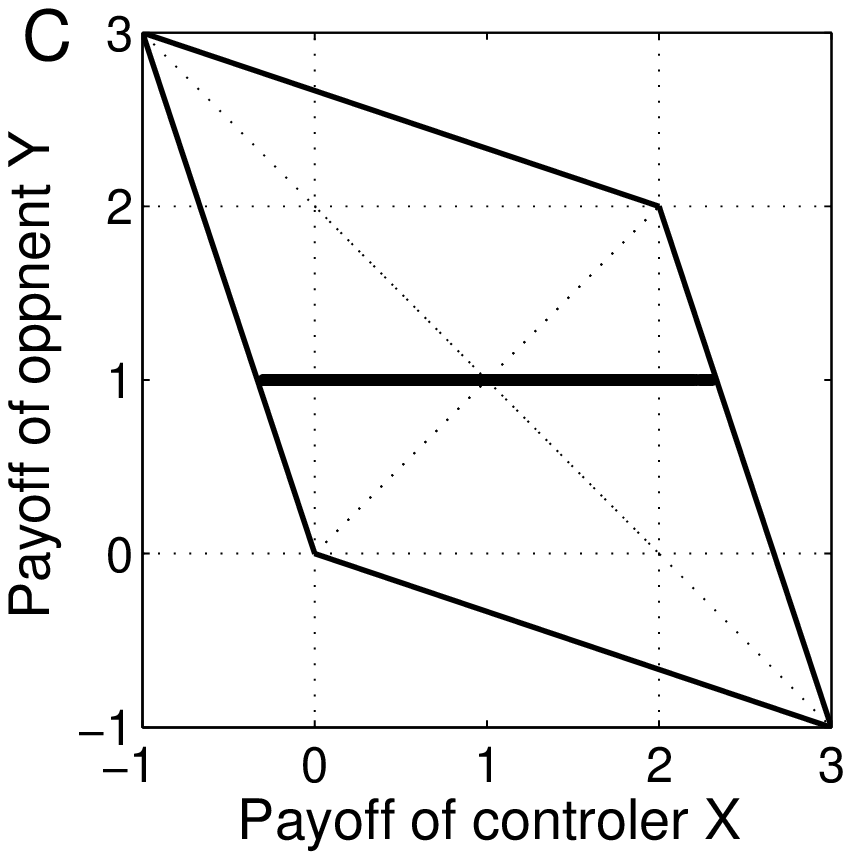}
    \caption{Control on maximum and minimum values of Y's possible payoffs. Each black dot is a possible payoff pair. In (A), (B) and (C), X's strategies are ${\bf p}=(1,0.51,1,0)$, ${\bf p}=(0.998,0.994,0.01, 0.0012)$ and ${\bf p}=(0.5, 0, 1, 0.5)$, respectively.}
\label{fig_probe_uw}
\end{figure}

%In (C), the strategy ${\bf p}=(0.99, 0.9775, 0.0275, 0.0082)$, compresses the payoff region of Y between $W=1.2$ and $U=0.8$;

It is straightforward that X can set $W$ and $U$ simultaneously and sandwich Y's expected payoff $s_Y$ into an intermediate region. She can do this by choosing a strategy $\bf p$ satisfying both eqs. (\ref{constrain_for_upper_bound_final}) and eqs. (\ref{constrain_for_lower_bound_final}). When $W$ and $U$ become closer to each other, the range of Y's possible payoff shrinks. In the extreme case, when controller X sets $W=U$, the region of Y's possible payoffs will be compressed into a line. At this point, $\bf p$ degenerates to an equalizer strategy, which has been discovered in \cite{boerlijst1997equal} and formally discussed in \cite{press2012iterated}.

In Figure 1 we show an example of how a payoff control on Y's maximum and minimum payoffs degenerates to a equalizer strategy. The convex hull is the space for the two players' payoff pairs $(s_X, s_Y)$. The x-axis and y-axis are the payoff values for player X and Y, respectively. In each subfigure, X uses a control strategy and Y's strategy is randomly sampled for 5000 times. We use a traditional prisoner's dilemma payoff matrix setting $(R,T,S,P)=(2,3,-1,0)$. Each black dot represents a possible payoff pair consisted of X's and Y's average payoffs under the fixed control strategy of X and a random strategy of Y. The upper and lower bounds of Y's possible payoffs are depicted by the red and blue lines, respectively. In 1(A), X's control strategy yields the maximum and minimum values $W=2$ and $U=0$ for Y; In 1(B), X sets $W=1.5$ and $U=0.5$ and the possible payoff region of Y shrinks; In 1(C), the general regional payoff control finally degenerates to an equalizer strategy, under which Y's payoff is pinned to a fixed value $W=U=1.0$. We can see that the equalizer/pinning strategy are special cases of control on the maximum and minimum values of opponent's payoffs.

%We can observe that, when $p_2$, $p_3$ and $p_4$ are determined, the upper bound of player Y's payoff is fixed. However, different value of $p_2$ will leads to different regions of player Y's possible payoff. A tough player X with small $p_2$ will make the payoff of player Y more uncertain, while a kind player X with large $p_2$ will make his own bottom line higher and make the region of player Y more close to the upper bound, offering player Y a higher probability to get a high payoff. Therefore the kindness and toughness of player Y can be depicted by using the payoff control strategy.

\begin{figure*}\label{fig_parterner_rival}
    \centering
	\includegraphics[width=0.19\textwidth]{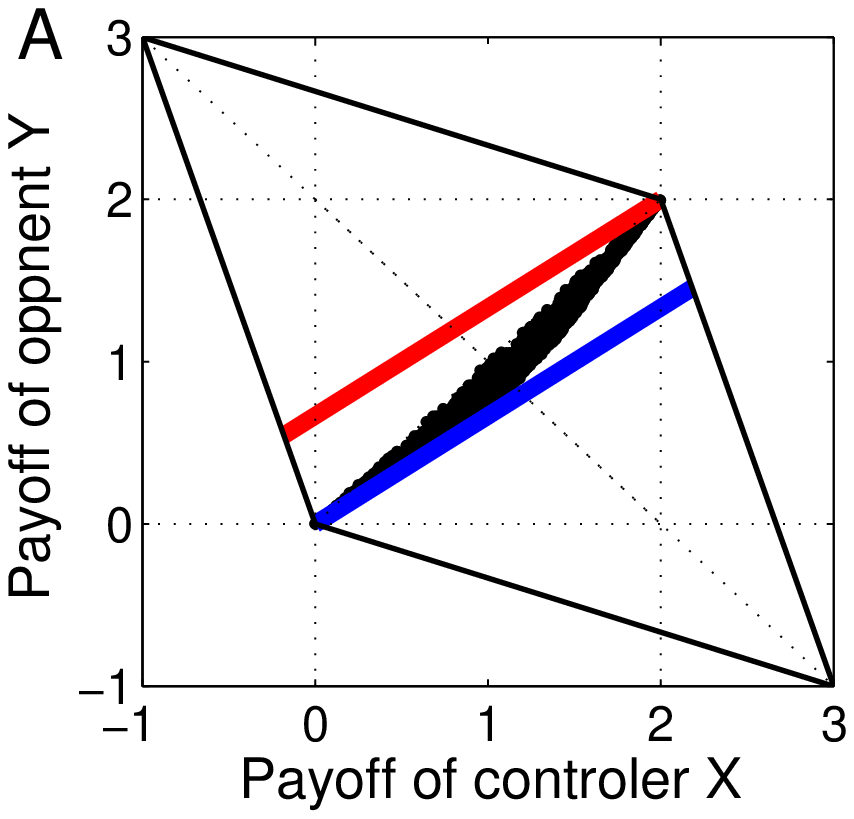}
    \includegraphics[width=0.19\textwidth]{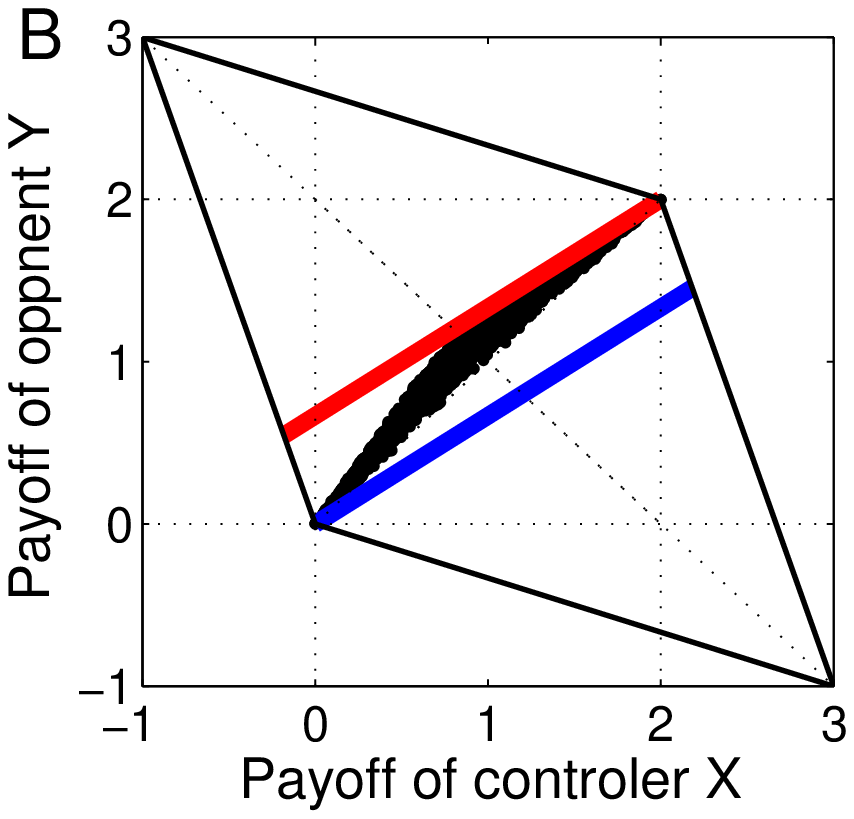}
    \includegraphics[width=0.19\textwidth]{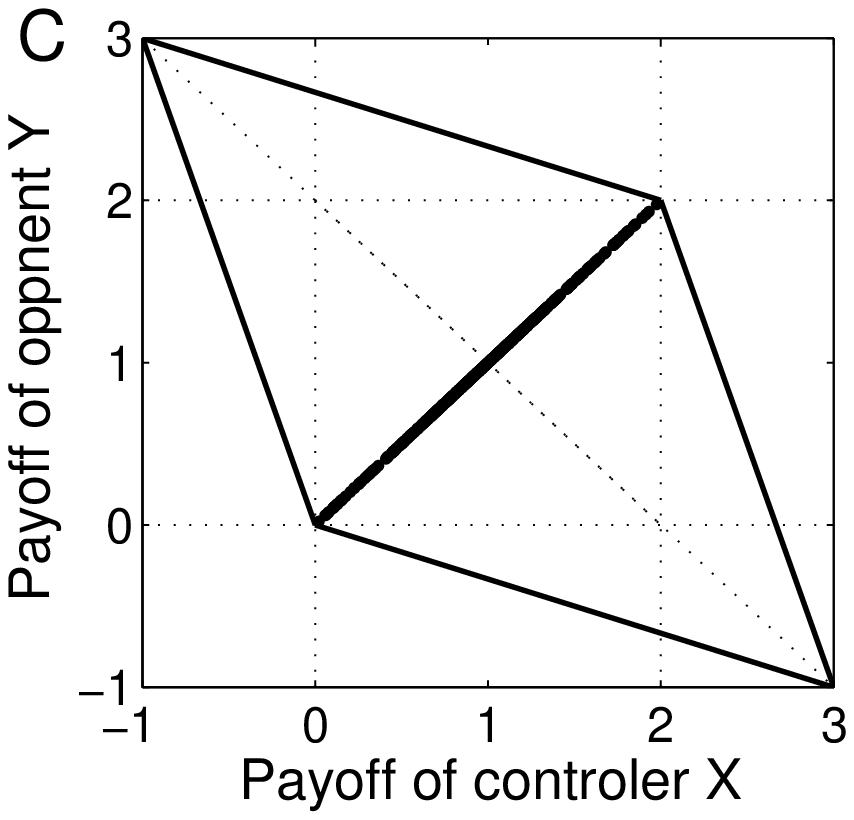}
   \includegraphics[width=0.19\textwidth]{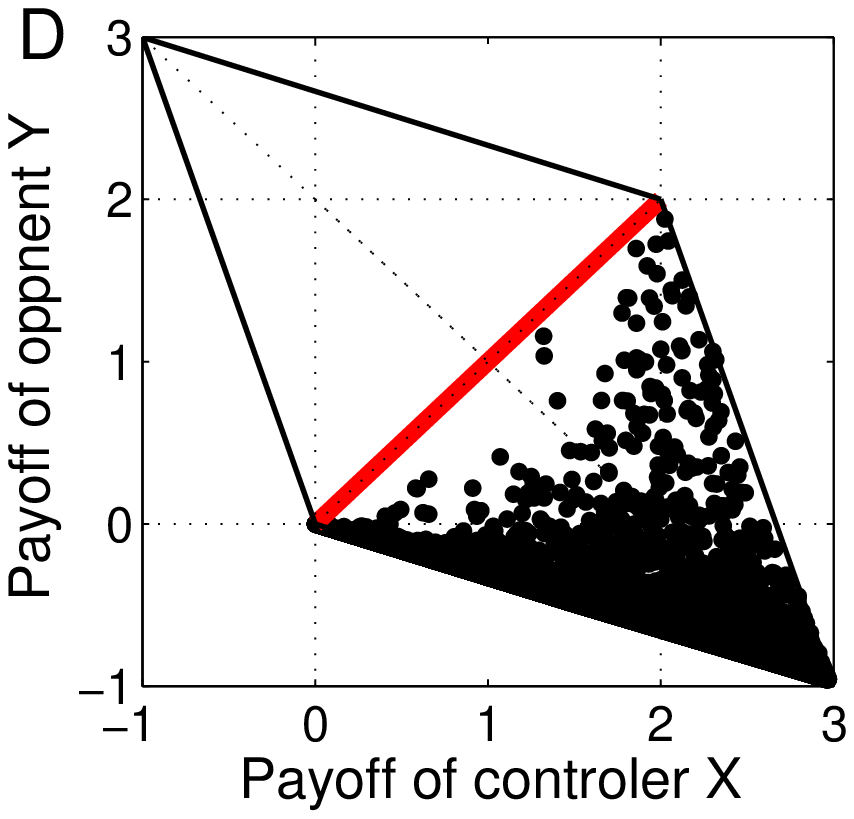}
    \includegraphics[width=0.19\textwidth]{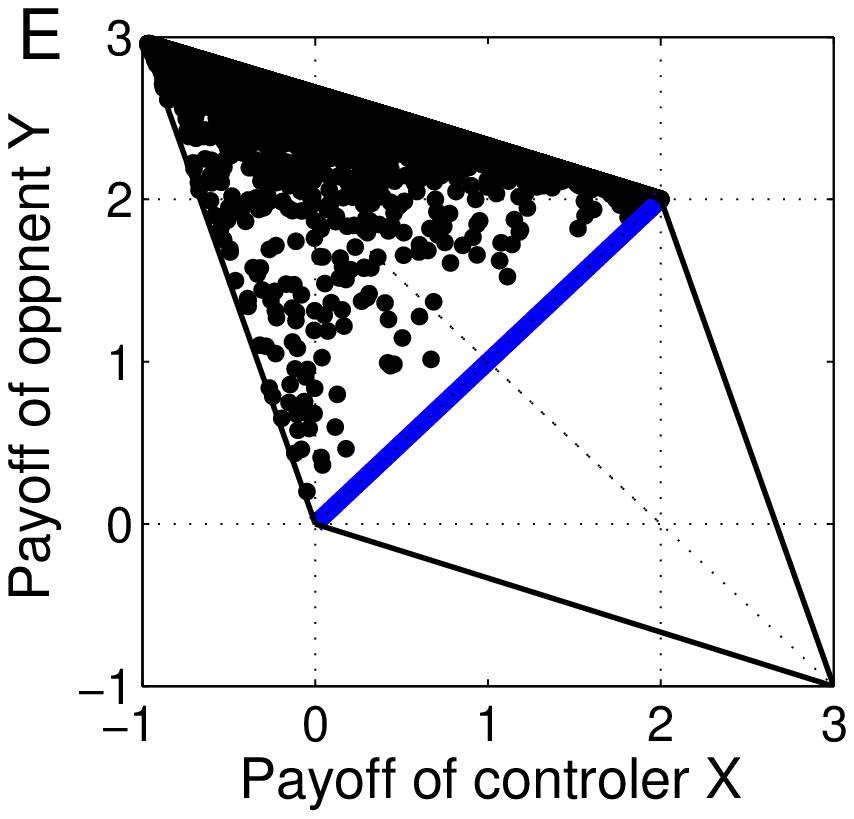}
    \caption{Control on the region of possible payoff pairs. Each black dot is a possible payoff pair. (A) is a normal winning/selfish control ${{\bf{p}}} = \left( {1,0.1,0.75,0} \right)$. (B) is a normal altruist control ${{\bf{p}}} = \left( {1,0.182,1,0} \right)$. (C) is a TFT-like strategy with $p_1=1, p_4=0$ and $p_2+p_3=1$. (D) is an extreme case of selfish control ${{\bf{p}}} = \left( {1,0,0.005,0} \right)$. (E) is an extreme case of altruist control ${{\bf{p}}} = \left( {1,0.995,1,0} \right)$. (A), (B), (C) and (D) are all cooperation-enforcing.}
\end{figure*}
\section{Control of Players' Payoff Relations and Cooperation Enforcement}
The above framework makes it possible to control the maximum and minimum values of the opponent's possible payoffs. In this section, we show that it is even possible for the controller X to confine the two players' possible payoff pairs in an arbitrary region of which the boundaries are characterized by linear functions. Under such a regional control, the game can be lead to a mutual-cooperation situation.

%Assume the controller X wants to establish a relationship between the two players' payoffs such that,
%\begin{eqnarray}\label{sX_minus_sY_original}
%{s_X} \ge \chi {s_Y} + \kappa.
%\end{eqnarray}
%This objective demands a linear upper bound of Y's payoff and ensure all payoff pairs under this upper bound ${s_X} - \chi {s_Y} - \kappa =0$.
Assume the controller X wants to establish a relation between the two players' payoffs such that the opponent always obtains less than a linear combination of what she earns:
\begin{eqnarray}\label{sX_minus_sY_original}
{s_Y} \le {1\over \chi}  {s_X} + \kappa ,
\end{eqnarray}
where $\chi \ge 1$ and $(1-{1\over \chi})R \ge \kappa \ge (1-{1\over \chi})P$. This objective claims a linear upper bound of Y's payoff and ensures that all possible payoff pairs are under it. Eq. (\ref{sX_minus_sY_original}) is equivalent to $\left( {{{\bf S}_X} - \chi {{\bf S}_Y}} + \chi\kappa\cdot{\bf 1} \right) \cdot {\bf{v}} \ge 0$, which further leads to $\left[ {\left( {R,S,T,P} \right) - \chi \left( {R,T,S,P} \right) + \chi\kappa\cdot{\bf 1}} \right] \cdot \left( {{v_1},{v_2},{v_3},{v_4}} \right) \ge 0$. Multiplying both side by $1-p_2$ and representing $v_2$ by using $v_1, v_3$ and $v_4$, we have: %$\varphi$
\begin{eqnarray}\label{sX_minus_sY_set}
{{\gamma _1}{v_1} + {\gamma _3}{v_3} + {\gamma _4}{v_4} \ge 0}  ,
\end{eqnarray}
where
\begin{eqnarray}\label{sX_minus_sY_gamma}
\left\{ \begin{array}{l}
 {\gamma _1} = \mu\left( { - 1 + {p_1}} \right) + \left[\left( {1 - \chi } \right)R + \chi\kappa \right]\left( {1 - {p_2}} \right), \\
 {\gamma _3} = \mu{p_3} + \left( {T - \chi S + \chi\kappa} \right)\left( {1 - {p_2}} \right), \\
 {\gamma _4} = \mu{p_4} + \left[\left( {1 - \chi } \right)P + \chi\kappa \right]\left( {1 - {p_2}} \right), \\
 \end{array} \right.
\end{eqnarray}
and $\mu=\left( {S - \chi T + \chi\kappa} \right)$.
%Making $\gamma _1, \gamma _2$ and $\gamma _3$ nonnegative simultaneously will ensure Eq. (\ref{sX_minus_sY_original}) holds. Similarly, if X wants to establish a payoff relation such that she always obtains less than a linear combination of Y's payoff, then her objective is
%\begin{eqnarray}\label{sX_minus_sY_original2}
%{s_X} \le \chi {s_Y} + \kappa,
%\end{eqnarray}
Making $\gamma _1, \gamma _3$ and $\gamma _4$ simultaneously nonnegative is sufficient to ensure that eq. (\ref{sX_minus_sY_original}) holds. Similarly, if X wants to establish a payoff relation such that Y always obtains more than a linear combination of what she earns, then her objective is
\begin{eqnarray}\label{sX_minus_sY_original2}
{s_Y} \ge {1\over \chi}  {s_X} + \kappa.
\end{eqnarray}
This indicates that X sets a linear lower bound of Y's payoff. Such an objective demands a payoff region above the line ${s_X} - \chi {s_Y} + \chi\kappa =0$. To realize this, she just needs to make $\gamma _1, \gamma _3$ and $\gamma _4$ in eqs. (\ref{sX_minus_sY_gamma}) nonpositive simultaneously. One necessary condition for a control strategy to exist is that the objective region should be \emph{feasible}, which means on the one hand, the objective region of the possible payoff pairs must intersect with the $(P,P)-(S,T)$ line, which is the left boundary of the payoffs in the iterated prisoner's dilemma, depicting the payoff pairs when Y unconditionally defects, i.e., ${\bf q}=(0,0,0,0)$; on the other hand, the payoff region must also terminates at some point on the $(R,R)-(T,S)$ line, which is the right boundary depicting the possible payoff pairs when Y unconditionally cooperates, i.e., ${\bf q}=(1,1,1,1)$.

Specifically, (1) if X controls by using objective function eq. (\ref{sX_minus_sY_original}) with $\chi\ge1$ and $\kappa=(1-{1\over \chi})P$, we have $ \left( {s_X} - P\right)\ge \chi\left({s_Y} - P \right)$. Any point in the objective region ensures that X's payoff difference to $P$ is \emph{at least} $\chi$ times of that of Y. Under such a strategy, X only concerns about herself winning the game regardless of the opponent's outcome. This feature well captures the selfishness in nature, therefore, we call such strategies the ``selfish control'' strategies. For example, in a game with $P=0$, if X sets eq. (\ref{sX_minus_sY_original}) with $\chi=1.5$ and $\kappa=0$, she always obtains more than $1.5$ times of what Y obtains. ${{\bf{p}}} = \left( {0.5,0.5,0.4,0} \right)$ is one of such selfish strategies. (2) If X's objective function is eq. (\ref{sX_minus_sY_original2}) with $\chi \ge 1$ and $\kappa=(1-{1\over \chi})R$, Y's payoff difference to $R$ will be \emph{at most} $1 \over \chi$ times of that of X, meaning that X is offering a benefit to Y at the expense of hurting her own benefit. Since in biological organisms, altruism can be defined as an individual performing an action which is at a cost to themselves but benefits another individual, we call this strategy the ``altruist control'' strategies. (3) However, if X controls with constraint $(1-{1\over \chi})R > \kappa > (1-{1\over \chi})P$, who can win the game is uncertain, since whether a payoff pair locates below the diagonal line $(P,P)-(R,R)$ still depends on Y's strategy. Thus we call them ``contingent control'' strategies.
%(4) However, any control objective with Eq. (\ref{sX_minus_sY_original}) and $\kappa<(1-{1\over \chi})P$, or control objective with Eq. (\ref{sX_minus_sY_original2}) and $\kappa>(1-{1\over \chi})R$ does not realize a feasible regional control strategy.

More generally, it is also possible for controller X to set up combinatorial objectives, such that there are multiple linear upper and/or lower bounds of Y's possible payoffs. She can do this by generalizing the constraint coefficients $\gamma$ to
\begin{eqnarray}\label{linear_constrain_extension}
{\bf{Gv}}' \ge {\bf{0}},
\end{eqnarray}
where ${\bf v}'=(v_1, v_3, v_4)$ and $\bf G$ is a coefficient matrix with each entry $\gamma _{ij}$ as the $j$-th coefficient from the $i$-th control objective. Following such a regularization, the complex payoff control problem is reduced to a formal linear programming. As long as $\bf G$ constitutes a feasible payoff region, the combinatorial control objective can be realized. Under this framework of regional control with multiple constraints, various shape of payoff regions can be generated. Especially, ZD strategies are extreme cases of regional control.  %Moreover, it is worth noting that, all the zero-determinant strategies can be viewed as regional payoff control with extreme constraints.

If each player has chosen a certain strategy and no one can benefit by changing his strategy while the other players keep theirs unchanged, then the current set of strategy choices and the corresponding payoffs constitute a Nash equilibrium. A strategy $p_N$ of player X is called a Nash strategy if player X can control the upper bound of player Y to $R$. Thus, under the general payoff control framework in eq.(\ref{linear_constrain_extension}), any strategy with $s_Y \le R$ as a tight constraint is a Nash strategy. According to the definition of Nash equilibrium, it is straightforward that any pair of Nash strategies constitute a Nash equilibrium. However, although a Nash strategy can induce a fixed upper bound $R$ of Y's payoff, it is possible for Y to choose an alternative strategy other than fully cooperating ($\bf q=1$), which still yields $R$ for herself but with the payoff for controller X smaller than $R$. This is why a Nash equilibrium is not necessarily a cooperative equilibrium. So how to select out an cooperation-enforcing Nash strategy is a problem. The controller can enforce cooperation by setting:
\begin{eqnarray}\label{CEC}
s_Y \le {1\over \chi } s_X + (1-{1\over \chi })R, {\kern 6pt}{\kern 1pt}{\kern 1pt}{\kern 1pt}{\kern 1pt}{\kern 1pt}  p_1=1,
\end{eqnarray}
where $\chi \ge 1$. Under such strategies of X, the only best response of Y is to fully cooperate, whereby both players finally receive payoffs $R$ which will lead the game to a win-win situation. We call them ``cooperation-enforcing control''.

Under the above framework, one can derive arbitrary regional control strategies, as long as the region has feasible linear boundaries. In Figure 2, we show several examples of regional control strategies. In 2(A) and 2(B), X sets same linear upper and lower bounds for the payoff region. The red lines are upper bounds with $\chi=1.5$ and $\kappa=-1$ and blue lines are lower bounds with $\chi=1.5$ and $\kappa=0$. In 2(A), X uses a selfish control where her payoff is always larger than that of Y. In 2(B), X uses an altruist control which always lets Y win. Both these two strategies are cooperation enforcing, leading the game to evolve to a mutual cooperation equilibrium. In 2(C), we shrink the controlled region to an extreme case by setting the upper and lower bounds identically with $\chi=1.0, \kappa=0$. The solution shows, as long as $p_1=1, p_4=0$ and $p_2+p_3=1$, the control strategies have similar effect as the traditional Tit-for-tat (TFT): equalizing the two players' expected payoffs. In 2(D) X uses an extreme case of selfish control while in 2(E) X uses an extreme case of altruist control. These two cases are also investigated as partnership and rival strategies in \cite{hilbe2015partners}.

%

%called the partnership and rival strategies in \cite{}  , which is a rival strategy, ensuring that Y's payoff is always less than that of X. In (E) X sets the lower bound by $\chi=1, \kappa=0$, then X's strategy is ${{\bf{p}}} = \left( {1,0.995,1,0} \right)$, which is an extreme case of partnership strategy, ensuring that X's payoff is always less than that of Y.

%Two good instantiations of cooperation-enforcing control strategies are the well-known Win-Stay and Loss-Shift (WSLS) and Tit-for-tat (TFT).

%\section{From Tit-for-Tat to a Nice Control}

\section{Control in Axelrod's Tournaments}

%Right up to today, it has been a fundamental challenge for many disciplines to understand how various strategies perform in multi-agent interactions, and what is the best strategy in repeated games and why it is the best. The most influential experiments for strategy evaluation is established by Robert Axelrod as his iterated prisoner's dilemma computer tournaments \cite{axelrod1981evolution}. Based on the payoff control framework, in this section, we derive a control strategy and simulate a tournament.

Right up to today, it has been a fundamental challenge for many disciplines to understand how various strategies perform in multi-agent interactions, what is the best strategy in repeated games and why it is the best. The most influential experiments for strategy evaluation are established by Robert Axelrod as his iterated prisoner's dilemma computer tournaments \cite{axelrod1981evolution}. Based on the payoff control framework, in this section, we derive several control strategies and simulate them in a tournament.

%the strategies include the well-known traditional strategies such as Tit-for-tat (TFT), Generous-TFT, Tit-for-two-tats (TF2T), hard-TFT, hard-TF2T, Win-stay-lose-shift (WSLS), Random, Probe, Probe2, Probe3, hard-Probe, all-C, all-D, Grim Trigger, hard-Majo, Calculator, Hard-Joss, and

%The simulated tournament is similar as in \cite{stewart2012extortion}. Besides the classic strategies, they implemented two ZD strategies: Extort\_2 with $s_X-P=2(s_Y-P)$ and Generous\_2 with $s_X-R=2(s_Y-R)$. The simulation shows the best performance is from Generous\_2, which is followed by GTFT and TFT. We add four regional control strategies into the tournament, including Altruist\_${\bf G}$, Selfish\_${\bf G}$, Altruist\_TFT and Selfish\_TFT. Here $\bf G$ denotes the control objectives in eq. (\ref{linear_constrain_extension}). We use a conventional IPD setting $(R,T,S,P)=(2,3,-1,0)$, and the Altruist\_${\bf G}$ is derived with respect to two objectives $ s_Y  \le \frac{1}{2}s_X  + 1$ and $s_Y  \ge \frac{3}{4}s_X  + \frac{1}{2}$, while the Selfish\_${\bf G}$ is derived with respect to $s_Y  \le \frac{3}{4}s_X $ and $s_Y  \ge \frac{1}{2}s_X $. Selfish\_TFT and Altruist\_TFT are using the same strategies as in figure 2(A) and 2(B), respectively. It is worth noting that Altruist\_${\bf G}$ is essentially an regional control expanded based on Generous\_2, while Selfish\_${\bf G}$ is expanded from Extort\_2. Both Selfish\_TFT and Altruist\_TFT can be viewed as expansion from original TFT.

The simulated tournament is similar as in \cite{stewart2012extortion} but uses a different IPD setting $(R,T,S,P)=(2,3,-1,0)$. Besides classic strategies, Stewart and Plotkin implemented two ZD strategies: Extort-2 with $s_X-P=2(s_Y-P)$ and Generous-2 with $s_X-R=2(s_Y-R)$. Their simulation shows the best performance is from Generous-2, which is followed by GTFT and TFT. We add four regional control strategies into the tournament, including Altruist\_${\bf G}$, Selfish\_${\bf G}$, Altruist\_TFT and Selfish\_TFT. Here $\bf G$ denotes the control objective matrix from eq. (\ref{linear_constrain_extension}). Altruist\_${\bf G}$ is derived with respect to two objectives $s_X - R \ge 2 (s_Y  - R)$ and $s_X - R \le \frac{4}{3}(s_Y - R) $, while the Selfish\_${\bf G}$ is derived with respect to $s_X - P  \le 2(s_Y - P) $ and $s_X - P  \ge \frac{4}{3} (s_Y - P) $. Selfish\_TFT and Altruist\_TFT are using the same strategies as in Figure 2(A) and 2(B), respectively. It is worth noting that Altruist\_${\bf G}$ is essentially a regional control expansion based on Generous-2, while Selfish\_${\bf G}$ is expanded from Extort-2. Both Selfish\_TFT and Altruist\_TFT can be viewed as expansions from original TFT.

Due to the inherent stochasticity of some strategies, the tournament is repeated $1000$ times. In a tournament, each strategy in the above set meets each other (including itself) in a perfect iterated prisoner's dilemma (IPD) game, and each IPD game has 200 stages. The average results are shown in Table 1. The shaded rows are for the control strategies derived under our framework. One can see that the Altruist\_${\bf G}$ has the best performance. It is better than Generous-2 and has much higher score than either TFT or GTFT. The Altruist\_TFT also performs better than TFT and GTFT. The Selfish\_TFT is a little tougher than TFT, although it has slightly higher number of wins. Analogously, using the above payoff control framework, one could also generate other regional control strategies which are better than the corresponding ZD strategies. Although no strategy is universally best in such tournaments, because a player's performance depends on the strategies of all her opponents as well as the environment of the game, the control framework still provides us a new perspective to formally quantify new nice strategies.

	\begin{table}
		\centering
		\begin{tabular}{llrr}
			\toprule
			\textbf{Name}  & ${\kern 24pt} \bf p$     & \multicolumn{1}{l}{\textbf{Score}} & \multicolumn{1}{l}{\textbf{Wins}} \\
			\midrule
			\rowcolor{black!20}ALTRUIST\_$\bf G$      & $(1, 2/15, 1, 1/3)$   & 1.66  & 0 \\
			GENEROUS-2     & $(1, 2/7, 1, 2/7)$    & 1.60  & 0 \\
			\rowcolor{black!20}ALTRUIST\_TFT    & $(1, 0.182, 1, 0)$    & 1.52  & 0 \\
			GTFT  & $(1, 2/3, 1, 2/3)$    & 1.46  & 0 \\
			TFT   & $(1,0,1,0)$           & 1.44  & 0 \\
			TF2T  &                       & 1.43  & 0 \\
			HARD\_TF2T &                       & 1.37  & 0 \\
			\rowcolor{black!20}    SELFISH\_TFT & $(1, 0.1, 0.75, 0)$   & 1.33  & 1 \\
			WSLS  & $(1,0,0,1)$           & 1.29  & 0 \\
			HARD\_PROBE &                       & 1.26  & 8 \\
			ALLC  & $(1, 1, 1, 1)$        & 1.18  & 0 \\
			PROBE2 &                       & 1.11  & 4 \\
			GRIM  & $(1, 0, 0, 0)$        & 1.08  & 4 \\
			HARD\_TFT &                       & 1.08  & 4 \\
			RANDOM & $(1/2, 1/2, 1/2, 1/2)$ & 0.92  & 10 \\
			HARD\_MAJO &                       & 0.91  & 13 \\
			PROBE &                       & 0.81  & 6 \\
			CALCULATOR &                       & 0.76  & 12 \\
			PROBE3 &                       & 0.72  & 10 \\
			HARD\_JOSS & $(0.9, 0, 1, 0)$      & 0.72  & 14 \\
			\rowcolor{black!20}      SELFISH\_$\bf G$ & $(5/7, 0, 13/15, 0)$  & 0.64  & 15 \\
			ALLD  & $(0, 0, 0, 0)$        & 0.45  & 20 \\
			EXTORT-2 & $(6/7, 1/2, 5/14, 0)$ & 0.45  & 19 \\
			\bottomrule
		\end{tabular}%
        \caption{Results of the IPD tournament}
		\label{tab:addlabel}%
	\end{table}%

\begin{figure*}\label{evolution}
    \centering
	\includegraphics[width=0.24\textwidth]{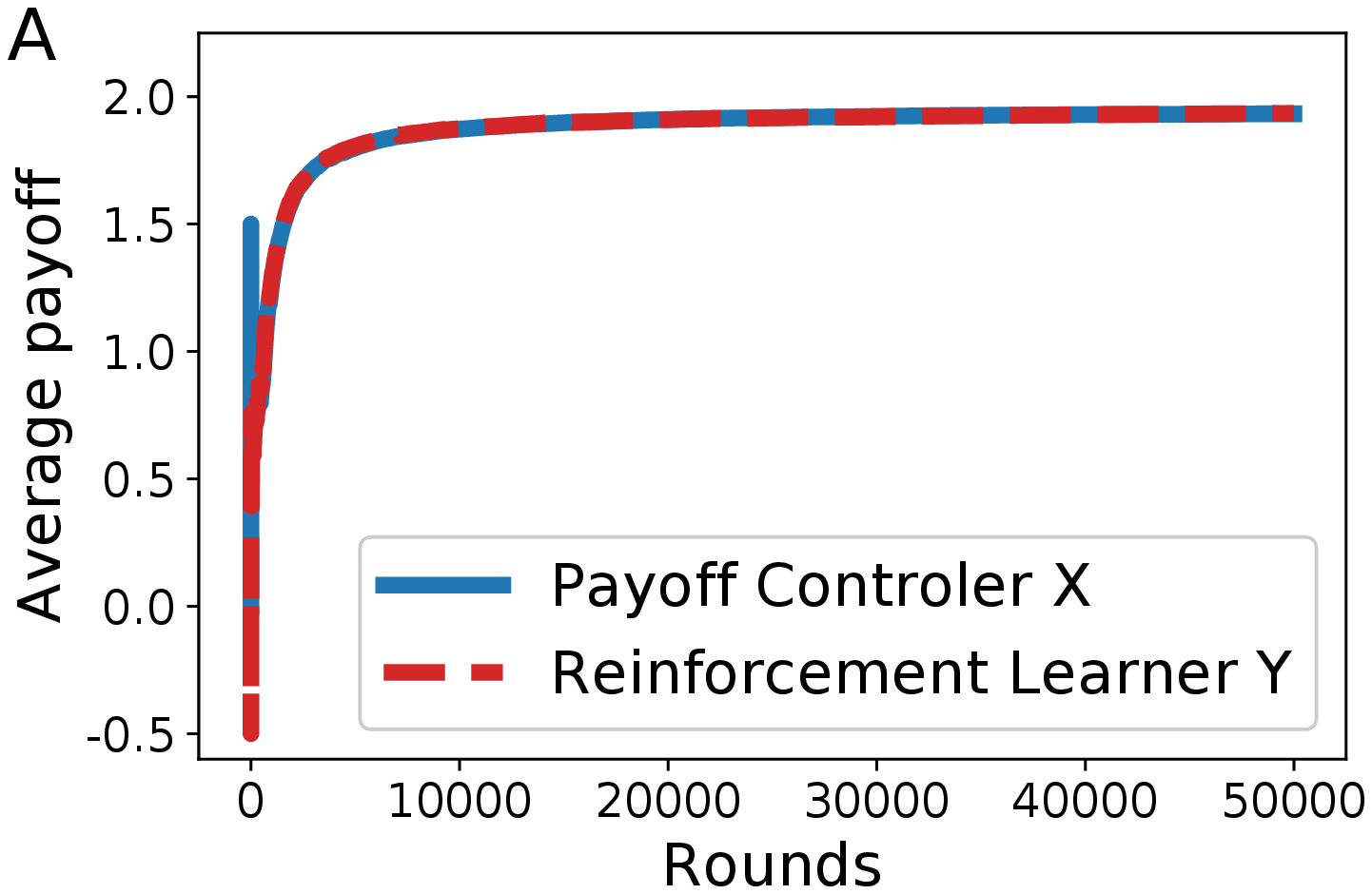}
    \includegraphics[width=0.24\textwidth]{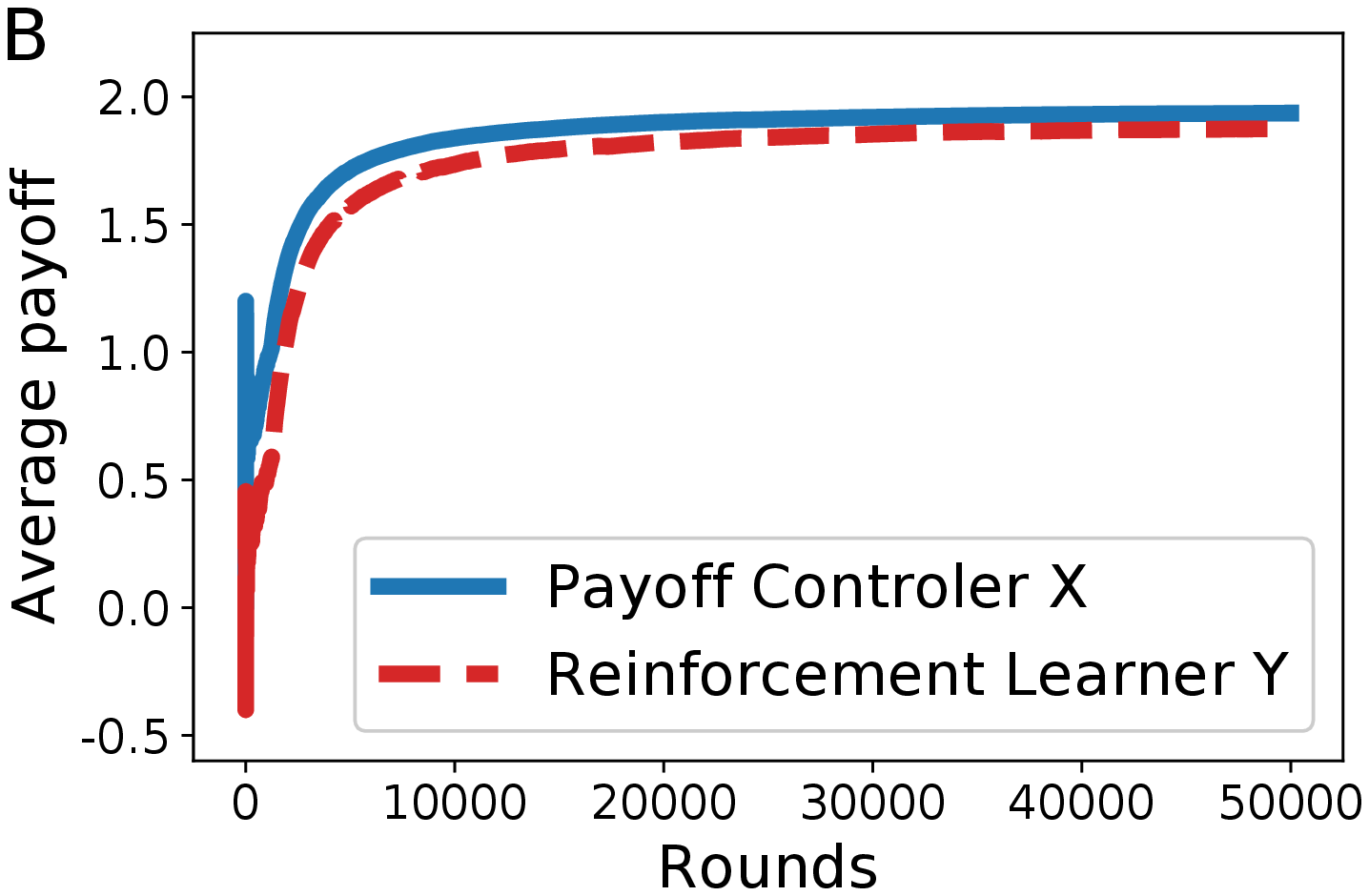}
    \includegraphics[width=0.24\textwidth]{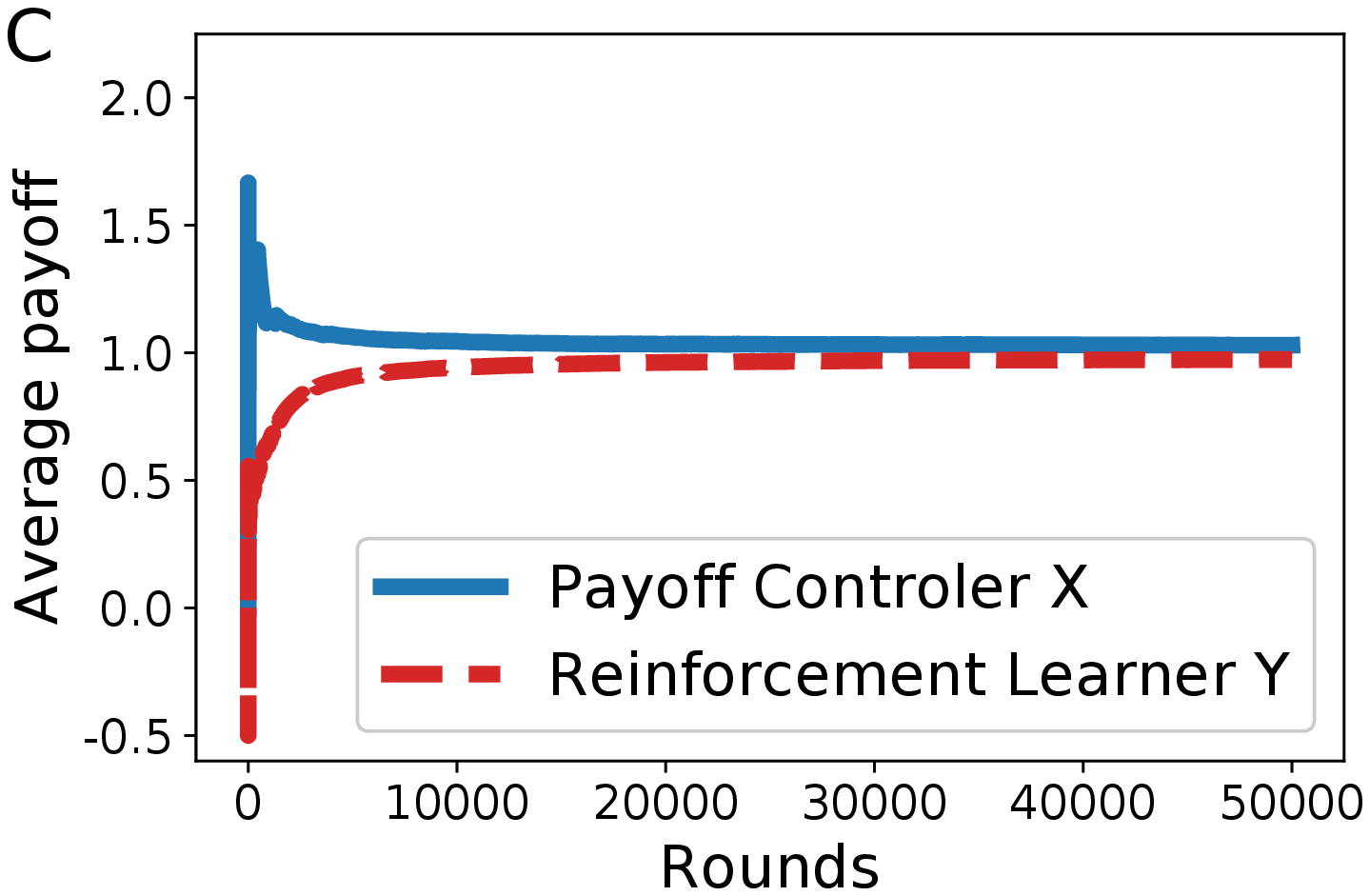}
    \includegraphics[width=0.24\textwidth]{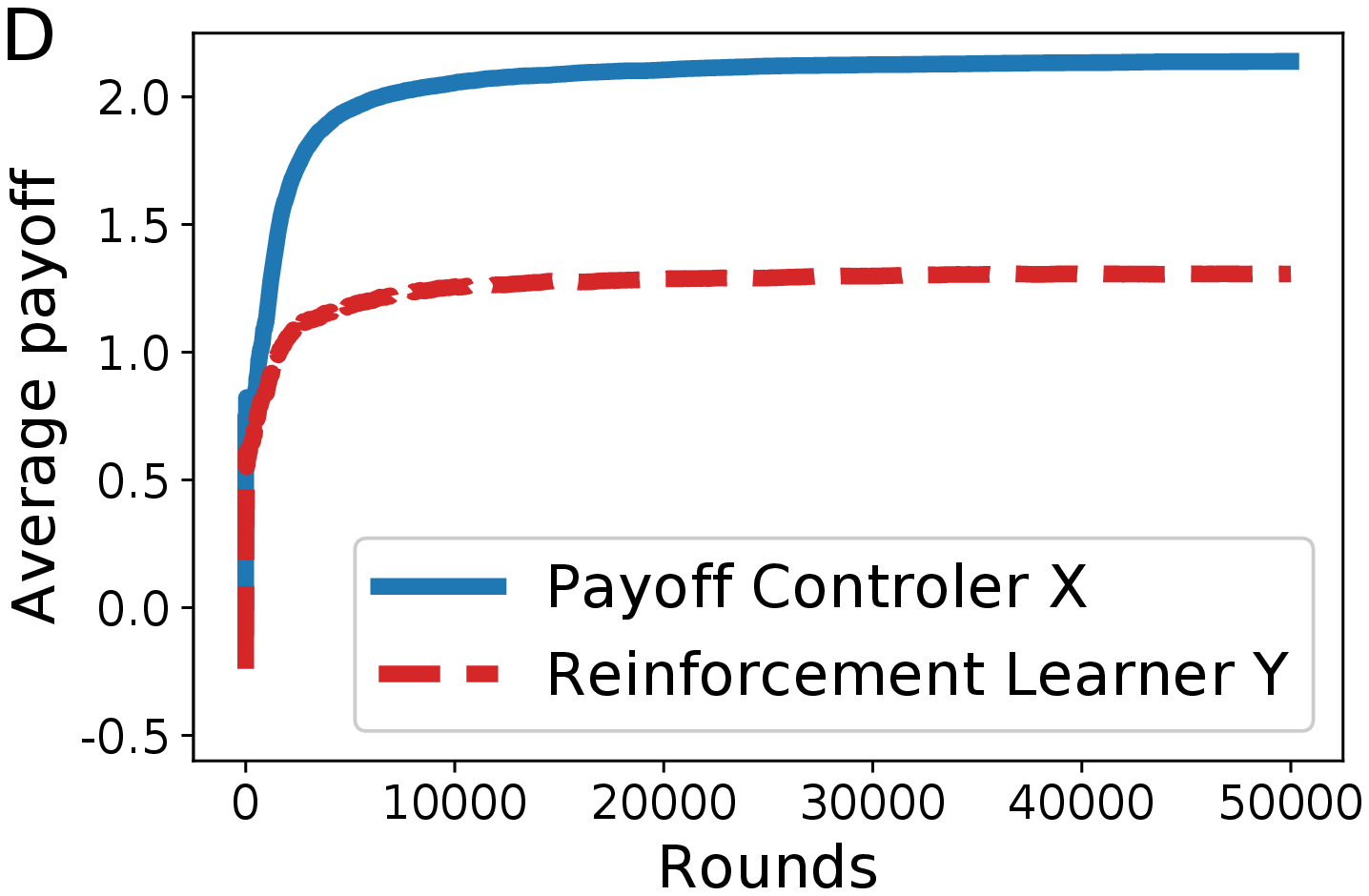}
    \caption{Control against human-like players. In each subfigure, Y uses a reinforcement learning strategy. In (A) X uses a TFT-like strategy ${\bf p}=(1, 0.2, 0.8, 0)$. In (B) X uses a selfish control ${\bf p}=(1, 0.1, 0.6, 0)$ which makes mutual cooperation $(s_X=R,s_Y=R)$ as the optimal outcome for Y. In (C) X uses selfish control ${\bf p}=(0.3, 0, 1, 0)$ which makes a payoff pair $(s_X=1,s_Y=1)$ as the optimal outcome for Y. In (D) X uses a selfish control ${\bf p}=(5/7, 0, 1, 0)$ which guarantees Y's payoff is much lower than that of X.}
\end{figure*}

In perfect environments, TFT has long been recognized as the most remarkable basic strategy. Starting with cooperation, it can constitute a Nash equilibrium strategy that enforces long-run cooperation. Nevertheless, TFT is not flawless. The first drawback of it, which is not apparent in perfect environment, is that if one of the two interacting TFT players faces the problem of trembling hand or imperfect observation, then a false defection will leads to a sequence of alternating cooperation and defection. Then the two players both receive a payoff much less than mutual cooperation. This indicates TFT is not a subgame perfect equilibrium strategy. Another weakness of TFT is a population of TFT players can be replaced by ALLC through random drift. Once ALLC has increased to some threshold, ALLD can invade the population. The reason why TFT is vulnerable to noise is that when confronting the opponent's unilateral defect ($CD$), a TFT player is too vengeful and will fully defect ($p_2=0$). The reason why TFT can be invaded by ALLC players is that it is not greedy at all, when it occasionally takes advantage of the opponent ($DC$), it completely stops defection and turns back to cooperation ($p_3=1$). To conquer these drawbacks, a nice strategy in a noisy environment should necessarily embody three features: (1) It should be cooperation-enforcing, i.e., its objective payoff region should have a tight upper bound $s_Y \le {1\over \chi } s_X + (1-{1\over \chi })R$ and $\chi \ge1$; (2) It should not be too vengeful, i.e., $p_2>0$, meaning its objective payoff region should not be too far from the $(S,T)$ point; (3) It should be somewhat greedy, i.e., $p_3<1$, meaning its objective payoff region should not be too far from the $(T,S)$ point.

\section{Control against Human-like Players}
In the real world, if a player is not aware of any nice strategies, he actually dynamically updates his stage action according to a learned history of the long-run game, and gradually evolves his own optimal plan for interacting with the opponent. In artificial intelligence, this learning and planning procedure is usually investigated by reinforcement learning models, which are state-of-the-art human like plays when agents are confronting with complex environment or strategic opponents. To try our best to understand the performance of the payoff control strategies in a real world, in this section, we simulate several repeated games between the payoff control players and the reinforcement learning players.

%In \cite{press2012iterated}, Press and Dyson provide a casual definition of an evolutionary player: facing X's fixed strategy, an evolutionary player Y is not aware of the theory of Markov strategies and improves his own payoff by adaptively adjusting his strategy following some dynamics. The authors then proposed a gradient-based optimization algorithm for Y. The convergence of the algorithm in ZD case is proved by Chen and Zinger in \cite{chen2014robustness}.

%Although it is easy and tempting to use a gradient-based optimization algorithm for Y, this kind of algorithm has two obvious shortcomings. First and the most important, the algorithm needs X's strategy as an input to compute the gradient, which makes the algorithm not practicable in the real world as Y often has no idea of X's strategy. We may approximate the gradient by iteratively playing the game with X but this approach is time consuming. Second, there is no more guarantee of convergence in general case. When X uses a generally memory-one strategy,  the convergence property of a gradient-based algorithm is unknown and the optimization process is easily getting into a local optimum in our experiments. For two reasons above, we must find a new optimization dynamics for Y.
%
%Here, we introduce a new optimization dynamics -- reinforcement learning.

Let X be the payoff controller who uses a payoff control strategy obtained beforehand, and let Y be the reinforcement learner who evolves his strategy/plan $\bf{q}$ according to the reinforcement learning dynamics. $\bf{q}$ is a mapping from the game history to the probabilities of selecting $a=C$. Y's objective is to find an optimal $\bf{q}^*$ which maximizes his stage payoff:
\begin{equation}\label{RL-policy}
{\bf{q}^{\rm{*}}}{\rm{ = }}\mathop {\arg \max }\limits_{\bf{q}}  \left\{ {\mathop {\lim }\limits_{n \to \infty } \frac{1}{n}\sum\limits_{t = 1}^n {\mathbb{E}_{\bf{q}}\left[ {s_Y^t} \right]} } \right\},
\end{equation}
where $s_Y^t$ is Y's realized stage payoff at time $t$ and $\mathbb{E}_{\bf{q}}$ is an expectation with respect to $\bf{q}$. Y's strategy $\bf{q}$ is updated according to the following average-reward value function:
\begin{equation}\label{RL-Q}
Q\left( {\omega,a} \right) \leftarrow \left( {1 - \alpha } \right) Q\left( {\omega,a} \right) + \alpha \left[ {{\bar r} + \mathop {\max }\limits_{a'} Q\left( {\omega',a'} \right)} \right]
\end{equation}
where $Q\left( {\omega,a} \right)$ is an evaluation value of player Y choosing action $a$ after stage game outcome $\omega$. ${\bar r}=r\left( {\omega,a,\omega'} \right)-r^*$ is difference between the instantiate reward $r$ and the estimated average reward $r^*$. The instantiation reward $r\left( {\omega,a,\omega'} \right)$ is induced by player Y taking action $a$ after outcome $\omega$ and transiting the game to a new outcome $\omega'$. $\alpha$ is a free variable for the learning rate. With $Q$'s values updated stage after stage, Y can improve his strategy dynamically \cite{gosavi2004reinforcement}.

We implement the above algorithm, simulate four repeated games and show the results in Figure 3. In each of these four games, player Y uses the reinforcement learning strategy described above. In 3(A) and 3(B), the strategies used by the controller X are both cooperation-enforcing. Under X's TFT-like control strategy in 3(A), X and Y always have almost the same average payoff; While under X's winning yet cooperation-enforcing strategy in 3(B), X dominates Y for a long time but the game finally converges to a mutual cooperation. In 3(C), X's objective is to set $s_X=s_Y=1$. We can see the game finally converges as X wishes. In 3(D), X uses a very tough selfish control, which means she can win Y a lot. In this situation, when the intelligent agent Y improves his own payoff step by step, he improves that of the controller even more. In a word, when playing against human-like reinforcement learning players, payoff control strategy players can lead the game to evolve to their objective outcomes.

%Moreover, those strategies which are cooperation-enforcing (such as strategies in 3(A) and 3(B)) are more meaningful in real-world multi-agent systems, since they can improve the systems' performance and maximize the social welfare.

\section{Conclusions}
We propose a general framework for controlling the linear boundaries of the region where the repeated game players' possible payoff pairs lie. By generating payoff control strategies under this framework, a single player can unilaterally set arbitrary boundaries on the two players' payoff relation and thereby realize her control objective, including limiting the maximum and minimum payoffs of the opponent, always winning the game, offering an altruist share to the opponent, enforcing the game to converge to a win-win outcome, and so on. The idea in this work is not limited to iterated prisoner's dilemma, it can be introduced into other two player repeated games and also can be generalized for repeated multi-player games, such as iterated public goods games. Future researches on the payoff control in repeated games with imperfect monitoring \cite{hao2015extortion}, with different memory sizes \cite{li2014effect} and researches investigating better winning or cooperation-enforcing strategies \cite{crandall2018cooperating,mathieu2015new} could be potentially inspired by this work. Furthermore, all the control strategies are based on the important premise that the player is with a theory of mind \cite{devaine2014theory}. Therefore, how to identify more cognitively complex human-like strategies in the context of the IPD, such as intention recognition \cite{han2012corpus}, is of great value for the future research.

\section*{Acknowledgments}
This work is partially supported by the National Natural Science Foundation of China (NNSFC) under Grant No. 71601029 and No. 61761146005. Kai Li also acknowledges the funding from Ant Financial Services Group.

%% The file named.bst is a bibliography style file for BibTeX 0.99c
\bibliographystyle{named}
\bibliography{ijcai18}

\end{document}